\documentclass[twocolumn,prl,eqsecnum,aps,floats]{revtex4}

\usepackage{dcolumn}
\usepackage{amsmath}
\usepackage{graphicx}
\usepackage{rotating}
\usepackage{amsfonts}
\usepackage{amssymb}
\usepackage[hang]{subfigure}

\usepackage{graphicx}
\usepackage{psfig,color}

\begin{document}
%
\author{Matthew Kleban and Raul Rabadan   }                                                                                                
\vskip 0.30cm                                                                 
                                                                           
\address{                                                                     
\centerline{Institute for Advanced Study, Einstein Drive, Princeton, NJ 08540}}


\def\MET{\mbox{${\hbox{$E$\kern-0.6em\lower-.1ex\hbox{/}}}_T$}}


\title {Collider Bounds on Pseudoscalars Coupling to Gauge Bosons}
\begin{abstract}
We bound the coupling of pseudo-scalar particles to ${\rm Tr}\, G\wedge G_{\rm QCD}$  using (the lack of) monojet plus missing $E_T$ events at the Tevatron, and estimate the bounds obtainable from LHC.  In addition, we revisit the bounds on the coupling to $F\wedge F_{\rm EM}$ from $e^+e^-$ collider events with single photon and missing $E_T$ final states.  This is especially interesting in light of the recent experimental results from the PVLAS collaboration, which we believe can be tested by data which will be available in the near future.
\end{abstract}
\maketitle

\section{Introduction}

In this note we use collider data to bound the couplings of pseudoscalar fields to gluons and photons. The processes we consider are $e^+e^- \rightarrow \gamma +$ \MET\ for $e^+e^-$ colliders, and $pp$ or $p\bar{p} \rightarrow$ single jet + \MET\ for hadron colliders. The signature is similar to that for gravitons flowing into extra dimensions, although they can be distinguished.  For instance, one  characteristic of the pseudo-scalar amplitudes is that the cross section is independent of center of mass energy at high energies.  This follows from dimensional analysis:  the interaction $g \phi F \wedge F$, where $\phi$ is the pseudoscalar field, has a coupling constant $g$ with dimensions of inverse mass.  Therefore $2\rightarrow2$ processes involving the production of a single $\phi$ particle will have a cross-section $d\sigma/d \Omega = g^2 f(s/t)$, where $f(s/t)$ is a function which depends only on the angle.   

Specifically, the couplings we will bound are the following \footnote{We have chosen the normalization to agree with standard coventions in the literature in the case $\phi$ is an axion.}:
\begin{equation}
\label{glue}
\frac{\alpha_s}{16 \pi f} \phi \  \ \epsilon_{\mu \nu \rho \sigma}  G^{\mu \nu}_a G^{\rho \sigma}_a
\end{equation}
for gluons, and 
\begin{equation}
\label{light}
\frac{g}{8} \phi  \  \  \epsilon_{\mu \nu \rho \sigma} F^{\mu \nu} F^{\rho \sigma} 
\end{equation}
for photons.  We will ignore any other possible couplings. 

There are a number of motivations, both theoretical and experimental, to be interested in such interactions.  Recently, the PVLAS collaboration has reported data consistent with the existence of a light, neutral, pseudoscalar particle coupled to the photon as above \cite{Zavattini}.  The favored mass range is $0.7 \ \ {\rm meV} < m_\phi < 2.0 \ \ {\rm meV}$, and the coupling $1.7 \ \ 10^{-6} \ \ {\rm GeV}^{-1} < g < 1.0 \ \ 10^{-5} \ \ {\rm GeV}^{-1}$.  These values appear to be in contradiction with astrophysical bounds coming from stellar dynamics and the lack of detection of such particles produced in the sun (see section 4. of \cite{Raffelt} for a recent review); however in some models these constraints might be relaxed \cite{Masso}.  In light of this, it is useful to have a direct bound on $g$ coming from particle physics data.  As we will see, an analysis of currently available data would come to the edge of this region of parameter space, and in the near future data will be available which will constrain it.

One bound on the coupling $g$ comes from decays of the $\Upsilon$ meson to photon plus missing energy.  The experiment CLEO \cite{Balest:1994ch} bounds the branching ratio $\Upsilon(1S) \rightarrow \phi \gamma < 1.3 \ \ 10^{-5}$, which translates (assuming the $\phi$ couples only to photons and not directly to quarks) into $g < 1.0 \ \ 10^{-3} \ \ {\rm GeV}^{-1}$.  A better bound $g < 5.5 \ \ 10^{-4} \ \ {\rm GeV}^{-1}$ was obtained using $e^+e^-$ collider data from ASP \cite{Masso:1995tw} \cite{Hearty:1989pq}.  As we will see, current collider data could improve this by nearly two orders of magnitude.

One particularly interesting pseudoscalar particle is the QCD axion, proposed as a solution to the strong CP problem \cite{axionbulk}, and which can also serve as a dark matter candidate.
For our purposes, there are two primary differences between axions and more general pseudoscalar particles with couplings (\ref{glue}) and (\ref{light}).  The first is the relation between the axion coupling and mass: $m_a \sim \Lambda_{QCD}^2/f_a$.  For bounds obtainable from colliders, the implication is that the mass of the axion can be effectively set to zero in the parameter range we are considering.  The second is the relation between $f_a$ and $g_a$:  $g_a = (\alpha C)/(2 \pi f_a)$, where $C$ is a model-dependent constant which is typically of order 1, but in special models can be significantly smaller.

\section{Summary of Results}

In this section we will summarize our estimates of the bounds on the pseudoscalar couplings which could be obtained with an analysis of collider data.  The details of our analysis can be found in the sections that follow.  We will give all bounds at the 95\% confidence level, and for experiments where the data have not yet been analyzed {\em we will quote bounds under the assumption that the results agree with the central value for the standard model background.}

Since we are looking for a missing energy signature, we will require that the pseudoscalar does not decay inside the detector.  Therefore the efficiency should include a factor of $\exp{(- L \Gamma/\gamma)}$, where $\Gamma$ is the inverse lifetime, $\gamma$ is $E_{\phi}/m_{\phi}$, and $L \sim 10$ m is the size of the detector.  The partial width for decay to photons is
\begin{eqnarray}
\Gamma_{\phi \rightarrow \gamma \, \gamma} &=&  {g^2 m_\phi^3 \over 64 \pi}.
\end{eqnarray}
For the case of non-zero $f$ and zero tree-level $g$, if the axion is too light to decay hadronically, the dominant decay mode will be to photons via a two-loop diagram involving gluons and charged hadronic matter.  Without assuming a specific model we can not determine precisely what the width will be; however generically this rate is small and the $\phi$ will not decay in the detector.  On the other hand, if $m_\phi$ is large compared to the mass of the pion the axion can decay hadronically and it is difficult to make a model-independent estimate of the width.  Therefore we will take the bound on the axion mass to be $m_\phi \sim m_{\pi^0}$.

In a particular model with calculable branching ratios, the bound at larger masses could be improved by looking for displaced vertices in the detector.  However, in what follows we will simply quote mass bounds using the requirement that the $\phi$ travel for a distance greater than ten meters, and assuming that the coupling saturates the stated bound.

Using the data from the D\O\ experiment at Fermilab:
\begin{equation}
f > 35 \  {\rm GeV}.
\end{equation}

A bound on the photon coupling $g$ was obtained in \cite{Masso:1995tw} for the process $e^+e^- \rightarrow \gamma +$ missing energy: $g \leq 5.5 \ 10^{-4} {\rm GeV}^{-1}$ for a pseudoscalar mass $m_{\phi} < 25 \ {\rm MeV}$\cite{Hearty:1989pq}.  This bound comes from the ASP detector at the Stanford Linear Accelerator Center storage ring PEP.

We have analyzed several past and future experiments to determine the bound which could be obtained assuming the data are consistent with the standard model backgrounds. 
Since the amplitudes are independent of energy, the bounds can be improved mainly by increasing the total luminosity.
An analysis of the combined data from LEP at ALEPH, OPAL, L3 and DELPHI would yield the more restrictive bound:  
\begin{equation}
g < 1.5 \ 10^{-4} \ \ {\rm GeV}^{-1}
\end{equation}
for $m_\phi < 65$ MeV.

As for future and ongoing experiments, we expect the Large Hadron Collider at CERN to improve the bound on $f$:
\begin{equation}
f_{\rm LHC} > 1300 \  {\rm GeV}.
\end{equation}

For the PEP-II $e^+e^-$ collider, current integrated luminosity gives 
\begin{equation}
g_{\rm PEPII} < 8.9 \ \ 10^{-6} \ \ {\rm GeV}^{-1}
\end{equation}
for $m_\phi < 0.12$ GeV. 

A similar bound can be obtained from KEKB $e^+e^-$ collider. With the current integrated luminosity the bound would be 
\begin{equation}
g_{\rm KEKB} < 8.2 \ \ 10^{-6} \ \ {\rm GeV}^{-1}
\end{equation}
for $m_\phi < 0.13$ GeV.  The expected total luminosity for KEKB is at least twice the current total, which would improve the bound to $g < 5.9 \ \ 10^{-6} {\rm GeV}^{-1}$.

Finally, for the Super KEKB upgrade to KEKB is expected to produce $10^7 \ \ {\rm pb}^{-1}$ per year.  After two years, this would improve the bound to 
\begin{equation}
g_{\rm KEKB} < 1.9 \ \ 10^{-6} \ \ {\rm GeV}^{-1},
\end{equation}
which would rule out most of the parameter space favored by PVLAS.

\begin{figure}
\begin{center}
\input{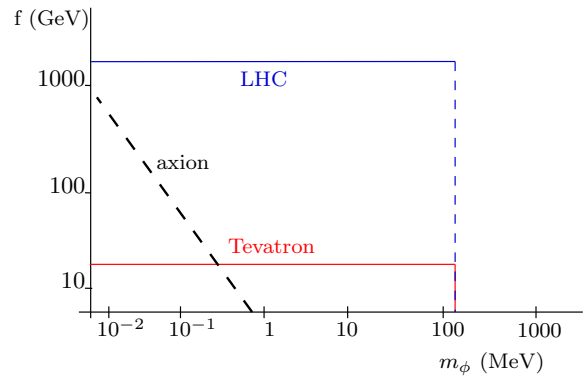}
\end{center}
\caption{95 \% confidence bounds on pseudoscalar-gluon couplings versus pseudoscalar mass obtainable for hadron colliders.  The testable regions for each experiment are below the line.}
\label{plot1} 
\end{figure}

\begin{figure}
\begin{center}
\input{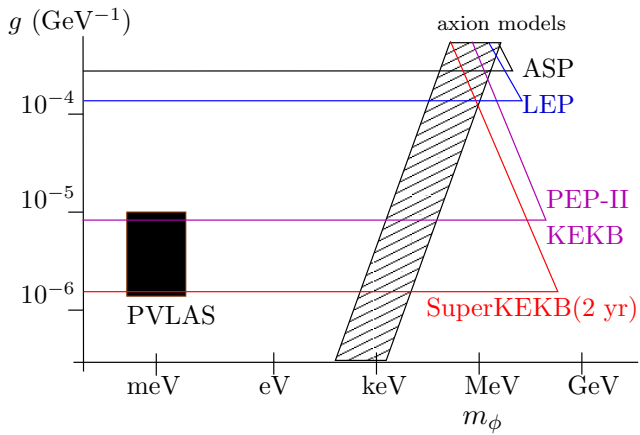}
\end{center}
\caption{Bounds on pseudoscalar-photon coupling versus pseudoscalar mass for $e+e^-$ colliders, the preferred region for the PVLAS results, and the band covered by typical QCD axion models.  The area above the lines could be tested with an analysis of collider data.}
\label{plot2} 
\end{figure}

In figures \ref{plot1} and \ref{plot2} we summarize the exclusion regions for these colliders.

\section{Bounds from hadron colliders}

\begin{figure}
\begin{center}
\input{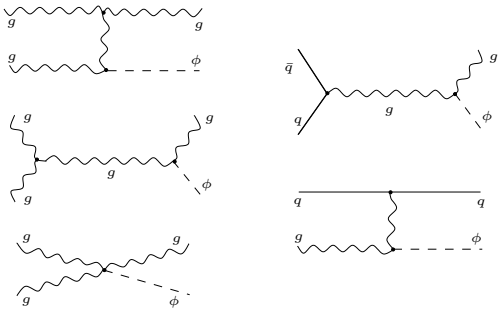}
\end{center}
\caption{Feynman diagrams for the process  $p\bar{p} \rightarrow \phi + {\rm jet}$.}
\label{QCD} 
\end{figure}

In this section we analyze the bounds on the pseudoscalar gluon coupling that can be obtained from $p\bar{p}$ annihilation to one jet plus \MET\ . The relevant subprocesses are represented in figure \ref{QCD}.

The data obtained by the D\O\  detector at the Tevatron between 1994 and 1996 corresponds to an integrated luminosity of 78.8 $\pm 3.9$ ${\rm pb}^{-1}$ with a center of mass energy $\sqrt{s} = 1.8$TeV.  The backgrounds for this process were analyzed in \cite{Abazov:2003gp}, in which the authors obtained bounds on a similar process in which the missing energy was lost into an extra dimension via Kaluza-Klein graviton emission.  For our analysis we will use the same cuts as \cite{Abazov:2003gp}: $\MET > 150 \ {\rm GeV}$, $E_T > 150 \ {\rm GeV}$, and pseudorapidity $|\eta_j| < 1.0$. The dominant background is the production of a $Z$ which decays to $\nu\bar{\nu}$, where the neutrinos carry the missing energy. 
The results of a numerical simulation using the software package CompHEP \cite{Pukhov:1999gg} are that the cross-section for monojet plus axion is $\sigma_{p \bar p \rightarrow \phi {\rm jet}} = 293 \ \ {\rm pb} \ \ ({\rm GeV}/f)^2$.

Using the error estimates of \cite{Abazov:2003gp} and requiring 95\% confidence, the bound obtained from this cross-section (assuming a detection efficiency of 1 given these cuts) is:
\begin{equation}
f > 35 \ {\rm GeV}.
\end{equation}

A similar analysis could be performed using the data from CDF \cite{Acosta:2003tz}. The luminosity considered in the analysis of \cite{Acosta:2003tz} is $84 pb^{-1}$. We expect the result to be similar. 

As for the Large Hadron Collider at CERN, we will take an integrated luminosity of $10^5$ pb$^{-1}$,  and select monojet events with $E_{T, jet} > 1$ TeV.  As we will see, given our estimates the signal will be limited by systematic error in the background rather than by statistics.  Again, the primary standard model background will be $pp \rightarrow \nu \bar{\nu}$ + $jet$.  At center of mass energy $\sqrt{s} = 14$ TeV, the cross-section for this background is $\sigma_{\nu \bar{\nu}} \sim 4 \ 10^{-3}$ pb, which corresponds to 400 events.  If we estimate 5\% error in the background and require 95 \% confidence ($N_{\phi} \sim 40$), we need $\sigma_{\phi} \sim 4 \ 10^{-4}$pb.  

Including the cut, this corresponds to a bound on $f$
\begin{equation}
f > 1300 \ {\rm GeV}.
\end{equation}

\section{Bounds from $e^+ e^-$ colliders}

\begin{figure}
\begin{center}
\input{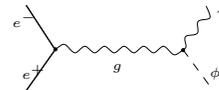}
\end{center}
\caption{Feynman diagram of the process  $e^+ e^- \rightarrow \gamma \phi$.}
\label{ee} 
\end{figure}

Consider the process $e^+ e^- \rightarrow \gamma \phi$ (see figure \ref{ee}). The  differential cross section is (in the limit $\sqrt{s} >> m_{e}$, but including the mass of the $\phi$ particle):
\begin{equation}
\left( \frac{d\sigma}{d\Omega} \right)_{CM} = \frac{\alpha g^2}{128 \pi} \frac{\left( s-m_\phi^2 \right) ^2}{s^2} \left(\cos^2{\theta} + 1\right) ,
\end{equation}
and the total cross section is $\sigma_T = \frac{\alpha g^2}{24}\frac{\left( s-m_\phi^2 \right) ^2}{s^2}  $.

One relevant fact is that the signal we are looking for has a two-body final state, with both particles effectively massless.  As a result, the photon always carries precisely half the center of mass energy.  On the other hand the standard model backgrounds have at least a three body final state, with zero phase space for the photon to carry this energy.  Furthermore, if the masses of some final state particles are significant, there is a further suppresion due to conservation of energy and momentum.  If the energy resolution of the detector were infinite, we could therefore cut out all the standard model background by requiring the photon energy to be very close to the beam energy.  In a real experiment, uncertainties in the photon energy will be relevant, and we take that into account in what follows.

As we mentioned above, this process was considered in \cite{Masso:1995tw}, using data from the ASP detector at SLAC. Improved bounds can be obtained using LEP data from ALEPH, OPAL, L3 and DELPHI. For instance, the ALEPH detector \cite{ALEPH} took data in a range of center of mass energies, between 188 and 209 GeV, with integrated luminosity 628 pb$^{-1}$.  The collaboration performed an analysis of single photon events, and found that the data are consistent with the predicted standard model background, which is dominated by $\gamma \nu \bar \nu$ in the final state.  However, for reasons discussed above a better bound on $g$ could be obtained with more stringent cuts on the energy of the photon.  The energy resolution of the ALEPH calorimeters is of order 3 GeV for a 100 GeV photon.  For example, for a beam energy of 94 GeV, if we require  $E_\gamma > 91$ GeV and $|\cos(\theta)| < .96$, $\sigma_{e \bar{e} \rightarrow \gamma \nu \bar{\nu}} = 2.7 \ 10^{-3} \ {\rm  pb}$, compared to $\sigma_{e \bar{e} \rightarrow \gamma \phi} = (g/10^{-5})^2 \ 1.2 \ 10^{-5} $ pb for the signal.  The expected number of background events is then 2, and requiring 95\% confidence \cite{Feldman}
 (4.72 events) would give a bound (assuming the results agree with the standard model)
\begin{equation}
g < 2.6 \ 10^{-4} \ {\rm GeV}^{-1},
\end{equation}
valid for $m_\phi < 48$ MeV.

Similar results can be obtained from  the other LEP experiments: L3 \cite{L3}, with  $189 < \sqrt{s} < 209$ and luminosity is 619 pb$^{-1}$; OPAL \cite{OPAL}, with $\sqrt{s} =$ 183 \ \ GeV,  and luminosity is 54 pb$^{-1}$ and DELPHI  \cite{DELPHI}, with $\sqrt{s} = 130 - 209 \ \ GeV $ and luminosity is 667 pb$^{-1}$.  Combining these gives roughly a factor of four in luminosity, but with these cuts we must require 6.8 events for 95\% confidence, and therefore the bound is $g < 1.5 \ 10^{-4} \ \ {\rm GeV}^{-1}$, for $m_\phi < 65$ MeV.

\vspace{0.5cm}
Better bounds can be obtained by higher luminosity colliders: PEP-II and KEKB. As explained above, requiring strict cuts on the energy of the photon (in the center of mass frame) greatly reduces the backgrounds due to many particles in the final state and particularly massive particles ($B \bar{B}$ processes, $K_L K_L \gamma$ processes where the $K_L$'s are interpreted as missing energy) and we do not expect them to be significant.  The energy resolution of the BABAR and BELLE detectors for a $~5$ GeV photon (in the center of mass frame) is around 1.5-2 \%. Therefore in the following analysis we will require that $E_{\gamma} > 0.985 E_{BEAM}$. 

The center of mass energy of both colliders is similar, as both are tuned to the $\Upsilon$(4S) resonance at $\sqrt{s} =10.6 \ \ GeV$.  However, due to the large width of the $\Upsilon$(4S) ($20$ MeV), there is no significant enhancement to the continuum cross section from the resonance.  Using the cut discussed above we obtain $\sigma_{e \bar{e} \rightarrow \gamma \phi} =  (g/10^{-5} {\rm GeV}^{-1})^2 \ 1.2 \ 10^{-5} \ {\rm pb} $, compared to   $\sigma_{e \bar{e} \rightarrow \gamma \nu \bar{\nu}} = 6.2 \ 10^{-7} \ {\rm pb}$. 

Taking a total integrated luminosity for PEPII of $312 \ \ fb^{-1}$, and  cuts in pseudorapidity $[-2,2] $ and photon energy, the number of Standard Model events is almost zero. Requiring three $\phi$ events gives a bound at 95\% confidence:
\begin{equation}
g < 8.9 \ 10^{-6} \ {\rm GeV}^{-1},
\end{equation}
for $m_\phi < 0.12 GeV$.

The integrated luminosity for KEKB at the time of writing is $ L = 5 \ 10^5$ pb$^{-1}$.   Imposing cuts as above in rapidity and in the photon energy the number of Standard Model events is again almost zero. Requiring three events gives a bound at 95\% confidence:
\begin{equation}
g < 8.2 \ 10^{-6} \ {\rm GeV}^{-1},
\end{equation}
for $m_\phi < 0.13 GeV$.

If luminosity increases to $ L = 1.0 \ 10^6$ pb$^{-1}$, almost 1 Standard Model events is expected. Requiring 4 events gives a bound at 95\% confidence:
\begin{equation}
g < 5.9 \ 10^{-6} \ {\rm GeV}^{-1},
\end{equation}
for $m_\phi < 0.16 GeV$.
 
For SuperKEKB the expected luminosity per year is $ L = 1.0 \ 10^7$ pb$^{-1}$. In a two year period, imposing similar cuts, the expected number of background events is 12.4. Requiring 8.8 additional events gives a bound at 95\% confidence:
\begin{equation}
g < 1.9 \ 10^{-6} \ {\rm GeV}^{-1},
\end{equation}
for $m_\phi < 0.28 GeV$.

\vspace{1cm}

\centerline{\bf Acknowledgements}

We would like to thank Kaustubh Agashe, Gregory Gabadadze, Ian Low, Satoshi Mishima, Hitoshi Murayama, Carlos Pe\~na Garay, Michael Peskin, Aaron Pierce, Massimo Porrati, Pierre Sikivie, Stewart Smith and Scott Thomas for useful discussions.  We would like to thank specially Concha Gonzalez Garcia for her insightful comments. In addition we would like to thank the speakers and organizers of the 2005 PITP school at the Institute for Advanced Study. R.R. is supported by DOE grant DE-FG02-90ER40542; M.K. is supported by NSF grant PHY-0070928.

\end{document}